\documentclass[aps,prstab,reprint,groupedaddress,amsmath,amssymb,floatfix, nofootinbib]{revtex4-2}

\usepackage{graphicx}% Include figure files
\usepackage{dcolumn}% Align table columns on decimal point
\usepackage{bm}% bold math
\usepackage{placeins}

\usepackage{comment}
\usepackage{geometry} % Used to adjust the document margins
\usepackage[pagebackref]{hyperref}
\renewcommand*\backref[1]{\ifx#1\relax \else ($\uparrow$ #1) \fi}
\usepackage{booktabs}

\usepackage{algorithm}
\usepackage[noend]{algpseudocode}

\usepackage{physics}
\usepackage{siunitx}
\DeclareSIUnit{\lightspeed}{\ensuremath{c}}
\DeclareSIUnit{\sigma}{\ensuremath{\sigma}}
\usepackage{array}
\usepackage{tabularx}

\usepackage{tikz}
\usetikzlibrary{arrows.meta, positioning, shapes.geometric, shapes.misc, fit, calc}

\begin{document}

\preprint{APS/123-QED}

\title{Surrogate-assisted beam distribution shaping at the CERN Proton Synchrotron and Super Proton Synchrotron} % Force line breaks with \\
\author{Matthias Remta}
\affiliation{CERN, Geneva 1211, Switzerland}
\affiliation{University of Vienna, Vienna 1090, Austria}
\author{Yann Dutheil}
\affiliation{CERN, Geneva 1211, Switzerland}
\author{Francesco Velotti}
\affiliation{CERN, Geneva 1211, Switzerland}

\date{\today}% It is always \today, today,

\begin{abstract}
The Super Proton Synchrotron at CERN delivers protons to the North Area experimental facility at \SI{400}{\giga\electronvolt\per\lightspeed}. The extracted beam is split twice vertically in the transfer line with Lambertson septa to serve three fixed-target experiments simultaneously. A significant fraction of the beam is lost during these splitting processes, driving radioactive activation of the Lambertson septa. This contribution investigates reducing the splitting losses by optimising the vertical beam distribution before extraction. The developed method relies on resonant excitation with the transverse damper, reshaping the Gaussian vertical profile into a uniform and ultimately bimodal distribution, with the excitation amplitude acting as a single continuous control knob. A loss reduction of up to \SI{23.5}{\percent} was measured. In simulations, the vertical amplitude detuning is established as a second knob that reduces the resulting beam envelope. This technique was verified with measurements at the CERN Proton Synchrotron. Surrogate models based on generative deep learning were developed for efficient real-time validation and Bayesian system identification.
\end{abstract}

\maketitle

\section{Introduction}\label{sect:introduction}
\begin{figure}
    \centering
    \includegraphics[width=1.0\columnwidth]{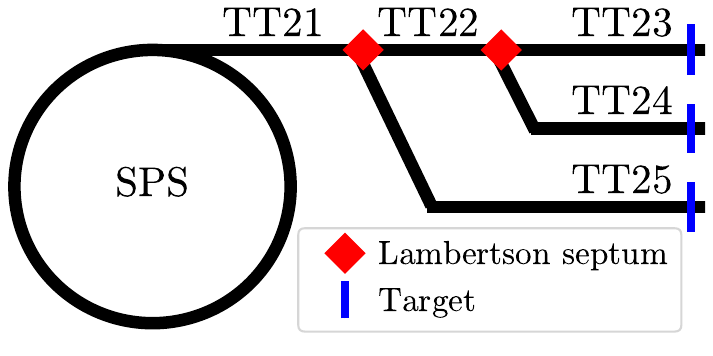}
    \caption{Schematic of the SPS and the primary transfer lines for the NA at CERN. The primary beam is split twice with Lambertson septa (red diamonds) to serve three fixed targets (vertical blue lines) simultaneously. The names of the transfer tunnels -- TT21/22/23/24/25 -- are shown above the respective section.}
    \label{fig:na_schematic}
\end{figure}

The North Area~(NA) at CERN is a versatile fixed-target facility integral to the laboratory's Physics Beyond Colliders programme~\cite{fernandez2025summaryreportphysicscolliders}. A broad mix of particles is produced at three primary targets and transported to the different experiments by secondary beam lines. The constant optics slow-extraction~\cite{PhysRevAccelBeams.22.101001} (COSE) scheme is employed in the Super Proton Synchrotron~(SPS) to provide spills of primary proton beams at \SI{400}{\giga\electronvolt\per\lightspeed} to the NA. This primary beam is vertically split twice with Lambertson septa~\cite{Evans:319349} in the transfer lines connecting the SPS with the NA, to serve the three fixed targets simultaneously. The layout of the primary transfer tunnels (TT21/22/23/24/25) is shown schematically in Fig.~\ref{fig:na_schematic}. By design, a significant fraction of particles interact with the septa blades during splitting. Therefore, collimators are installed upstream to shadow and protect the blades. Across the two splitting stages, up to \SI{8}{\percent} of the beam is lost (see Appendix~\ref{app:sharing}). These losses radioactively activate beam-line elements, shorten component lifetimes, constrain maintenance work, and are among the limiting factors for the annual number of protons on target delivered to the NA~\cite{koto2022searchesnewphysicshighintensity}.

\begin{figure}
    \centering
    \includegraphics[width=1.0\columnwidth]{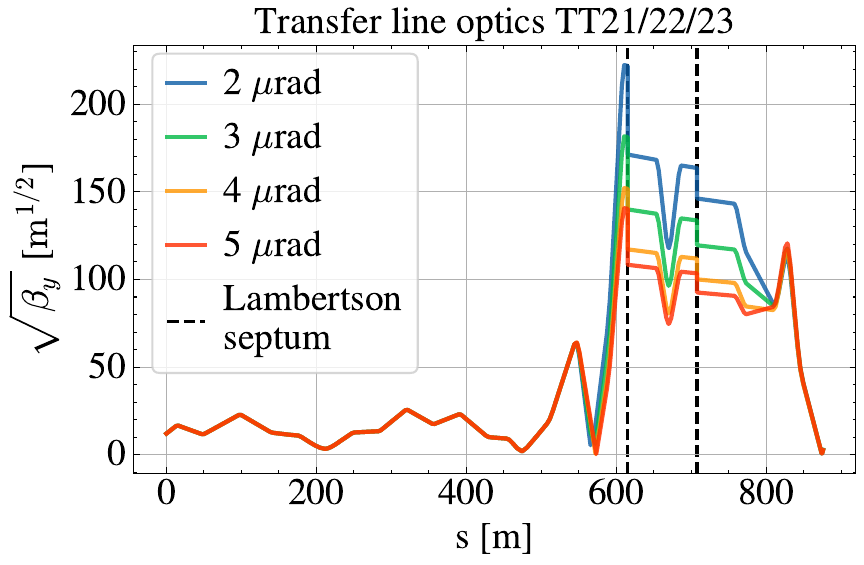}
    \caption{The square roots of the vertical \(\beta\)-functions along the transfer line TT21/22/23 are shown for four different configurations of linear optics. These optics are developed to maintain a fixed vertical beam size at the Lambertson septa for different \(\epsilon_y^N\) from \SIrange{2}{5}{\milli\meter\milli\radian} delivered by the SPS.}  
    \label{fig:tt20_optics}
\end{figure}

\begin{figure*}
    \centering
    \includegraphics[width=1.0\textwidth]{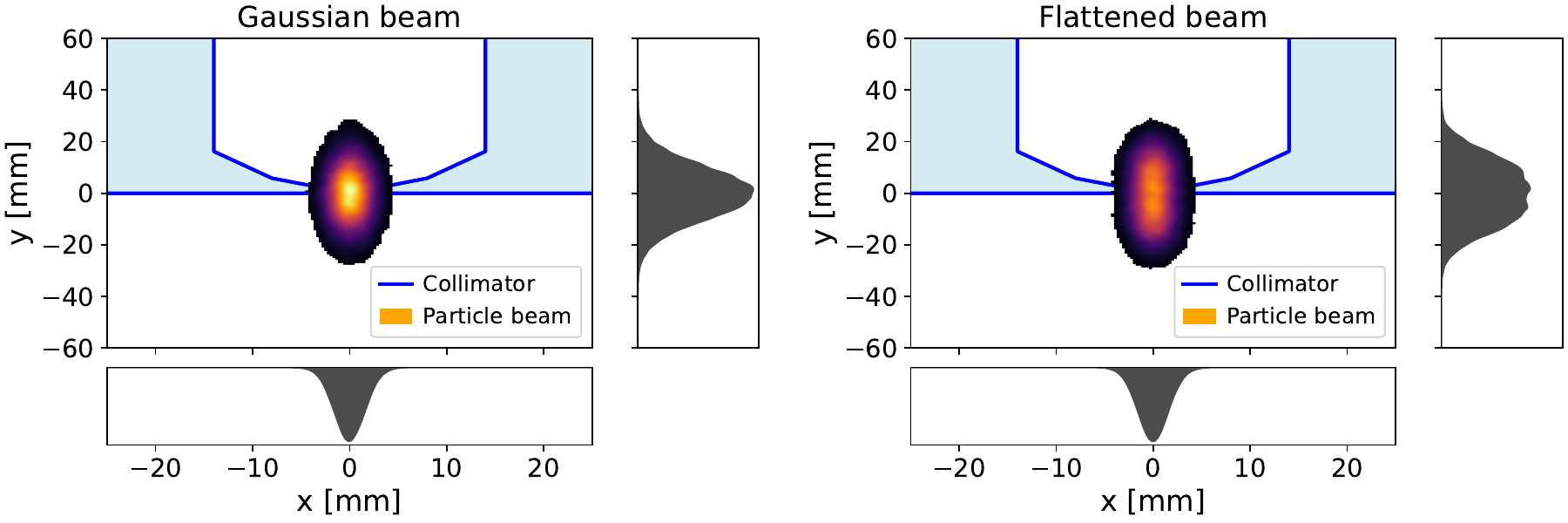}
    \caption{A simplified geometry of the collimators upstream of the Lambertson septa is shown in both panels (blue). Left: two-dimensional transverse density of a Gaussian particle beam and the two one-dimensional projections. Note that the beam is slow-extracted in multiple momentum slices and the horizontal Gaussian distribution is a simplification of the true profile. The intersection of the particle beam with the collimator geometry is lost during splitting. Right: the same projections for a vertically flattened particle beam. For a fair visual comparison, the 99.7\% envelope of the flattened beam was scaled to the \SI{3}{\sigma} envelope of the Gaussian beam.}  
    \label{fig:septum_geometry}
\end{figure*}

The splitting losses are negatively correlated with the vertical beam size at the septa, so they can be reduced by maximising the beam size through the linear optics. Four optics configurations are in place to maintain large vertical beam sizes at the septa while allowing different normalised vertical emittances \(\epsilon_y^N\) from \SIrange{2}{5}{\micro\radian} in the SPS (see Fig.~\ref{fig:tt20_optics}). This linear loss-mitigation strategy is ultimately limited by the available aperture in the transfer lines. Further loss reductions require altering the vertical beam distribution from a Gaussian towards a uniform or bimodal shape (see Fig.~\ref{fig:septum_geometry}). Such a transformation can be achieved with non-linear focusing~\cite{164748, PhysRevSTAB.10.104001}; however, the NA transfer lines are presently not equipped with non-linear magnets. Instead, this contribution investigates shaping the vertical beam profile in the SPS before extraction, using the vertical transverse damper to excite the beam until a hollow vertical distribution emerges. As this process only causes a moderate increase in beam envelope, it enables loss reductions beyond the linear method. The increase in beam envelope can be further reduced by inducing vertical amplitude detuning with octupoles, which was studied in simulations for the SPS and experimentally verified in the CERN Proton Synchrotron (PS).

Driving the transverse particle motion with an external excitation signal is the fundamental principle of radio-frequency knock-out (RFKO)~\cite{HIRAMOTO1992154, TOMIZAWA1993399}. In the RFKO context, the formation of hollow beams has been observed in simulations~\cite{CortesGarcia:2024sfh}, and a suitably configured amplitude detuning combined with an excitation signal can lead to auto-resonance, driving the beam to very high transverse amplitudes~\cite{s85n-289k}. The driven particle motion can be described with normal forms~\cite{PhysRevSTAB.5.054001} and a generalisation of Hill's equation~\cite{PhysRevSTAB.11.084002}. Various manipulations of the beam distribution with excitation signals have been theorised and simulated~\cite{PhysRevE.106.034204, PhysRevAccelBeams.26.024001, PhysRevAccelBeams.28.014001}, and the impact of pulsed excitation on the emittance in the LHC has been studied in Ref.~\cite{PhysRevAccelBeams.24.021001}. We extend this literature by (i) directly measuring the profiles generated with different excitation signals and levels of amplitude detuning, (ii) comparing them with single-particle simulations, (iii) applying this beam-shaping technique for loss reduction in a downstream transfer line, (iv) developing deep-learning surrogate models as a faster alternative to conventional particle tracking and (v) applying these models to real-time verification and system identification. The remainder of this article is organised as follows. Section~\ref{sect:sps} describes the beam shaping studies in the SPS and the surrogate model developed for them. Section~\ref{sect:ps} discusses the interplay with amplitude detuning in the PS and develops an effective model via Bayesian system identification. Finally, Sec.~\ref{sect:discussion} provides a discussion and Sec.~\ref{sect:conclusion} concludes the paper.

\section{Particle beam profile shaping in the SPS}\label{sect:sps}
\subsection{Setup and simulations}
The proposed beam-shaping method is applied at flat-top, at the start of slow extraction. The SPS uses the COSE~\cite{PhysRevAccelBeams.22.101001} scheme, in which the horizontal tune is set close to a third-integer resonance and strong sextupoles create unstable regions in phase-space. The radio-frequency cavities perform a voltage jump and are then switched off, which creates and maintains a large momentum spread. Through a normalised horizontal chromaticity of \(-1\), this momentum spread translates into a large tune spread. The currents in all magnets are then ramped simultaneously, driving successive momentum slices into resonance to produce a spill of \SI{1.2}{\second} or \SI{4.8}{\second}. For simplicity, we will also refer to these two spill lengths as short spill and long spill.
We reproduced this configuration in the \texttt{Xsuite}~\cite{Iadarola:2023fuk} model of the SPS and added the excitation signal on top. The linear tunes are \(Q_x = 26.62\) and \(Q_y = 26.58\), and the four extraction sextupoles are set to \(k_2 = \SI[per-mode=power]{-0.11992}{\per\cubic\meter}\). The excitation chirp is a linear frequency sweep parametrised by its amplitude \(A_{\text{ex}}\) (in \si{\micro\radian}), its frequency range \((q_{\text{ex, final}} - q_{\text{ex, initial}})\) in units of \(f/f_{\text{rev}}\), and its length \(n\) in turns. In addition to these three free parameters, the vertical normalised emittance \(\epsilon_y^N\) and the maximum momentum spread \(\delta_{\text{max}}\) can differ between operational scenarios and are treated as additional parameters in the simulations. The bounds imposed on these parameters for the simulations are stated in Tab.~\ref{tab:param_bounds}. We used Sobol sampling~\cite{OWEN1998466} to select \num{595} points in this five-dimensional parameter space. For each parameter combination, \num{1e4} particles from a matched initial distribution were tracked and the final phase-space coordinates recorded.

\begin{table}
\caption{Bounds of the parameter space for the SPS simulations. 595 Sobol points were sampled from this space.}
\label{tab:param_bounds}
\begin{ruledtabular}
\begin{tabular}{ll}
Parameter & Range \\
\midrule
$A_{\text{ex}}$ & $[\SI{0}{\micro\radian},\ \SI{0.5}{\micro\radian}]$ \\
$q_{\text{ex,f}}-q_{\text{ex,i}}$ & $[-0.04,\ 0.04]$~$f/f_{\text{rev}}$ \\
$n$ & $[128,\ 32768]~\text{turns}$ \\
$\epsilon_y^N$ & $[\SI{2}{\milli\meter\milli\radian},\ \SI{5}{\milli\meter\milli\radian}]$ \\
$\delta_{\text{max}}$ & $[\num{8e-4},\ \num{1.5e-3}]$ \\
\end{tabular}
\end{ruledtabular}
\end{table}

\subsection{Surrogate model}\label{sect:surrogate_model}
A generative surrogate model of the SPS was developed based on Conditional Flow Matching (CFM)~\cite{lipman2023}. Given a set of machine parameters, the model maps a six-dimensional standard-normal input to the corresponding final beam distribution in phase space by integrating a learned vector field. Conditioning the vector field on the machine parameters allows a single model to cover the full parameter range of interest~\cite{remta2026flowbasedsurrogatemodelsparticle}. Of the \num{595} SPS simulations \num{100} were used for training and the remaining \num{495} for testing. At the time the model was trained, only a small set of simulations was available. Due to additional simulations conducted at a later time, the fraction of test data is unusually large. For evaluation, we calculated the squared maximum mean discrepancy (\(\text{MMD}^2\))~\cite{gretton2012} between the particle ensembles produced by the model and the ground truth on the test set (see Fig.~\ref{fig:mmd_violin}). After this successful performance evaluation, we used the surrogate for real-time verification during measurements on the SPS and to generate larger ensembles of \num{2e5} particles from the final beam distribution for analysing losses in the transfer lines.

\subsection{Measurements}\label{sect:measurements}
\begin{figure*}
    \centering
    \includegraphics[width=1.0\textwidth]{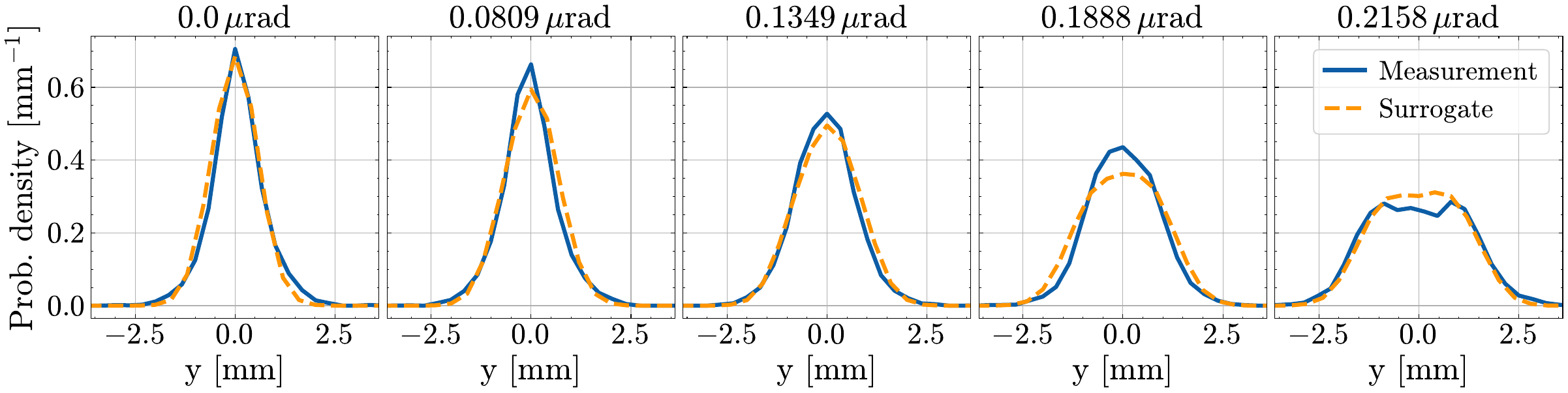}
    \caption{Wire-scanner measurements (blue) of the vertical beam profile in the SPS at different transverse-damper amplitudes \(A_{\text{ex}}\), as indicated in each panel title. The deformation of the beam profile grows with \(A_{\text{ex}}\). The surrogate model (orange dashed) predicts the same trend, demonstrating consistency between simulations and machine.}
    \label{fig:sps_ws_profiles}
\end{figure*}

During our measurement campaigns, we used both types of slow-extraction cycles, short spill and long spill. For profile measurements in the SPS ring the beam was kept at low intensity and disposed at the internal dump, to allow safe operation of the wirescanners and to avoid interrupting data-taking by the NA experiments. The transverse damper was configured with initial and final exciter tunes of \(0.41\) and \(0.45\), corresponding to \(1 - q_v \mp 0.02\) with \(q_v\) the fractional part of the vertical tune. The chirp thus covers the entire vertical tune spread of the beam. The chirp length was \num{2048} turns, corresponding to \SI{0.05}{\second}. The amplitude \(A_{\text{ex}}\) is controlled by a dimensionless gain between 0 and 1. Its equivalent kick amplitude in \si{\micro\radian} was extrapolated from damping-performance measurements at injection energy. We ramped \(A_{\text{ex}}\) stepwise up to \SI{0.2158}{\micro\radian}. The resulting vertical beam profiles are shown in Fig.~\ref{fig:sps_ws_profiles}. With increasing \(A_{\text{ex}}\) the Gaussian profile is more and more distorted and a bimodal distribution starts to emerge.

\begin{figure*}
    \centering
    \includegraphics[width=1.0\textwidth]{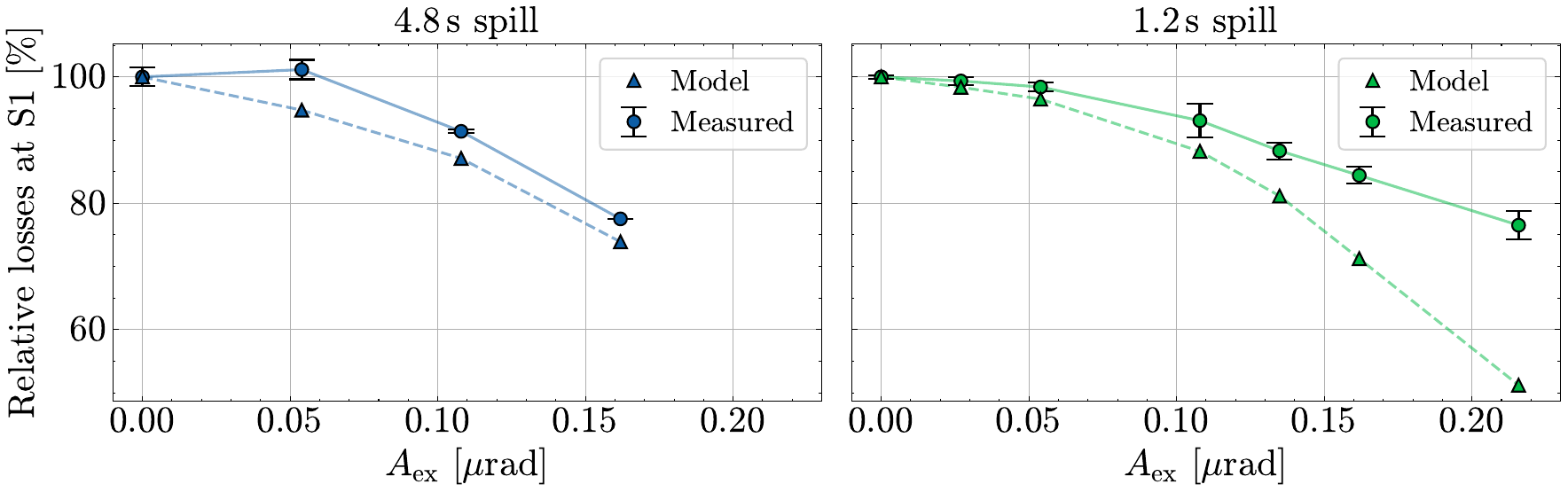}
    \caption{Losses at the first Lambertson septum versus the transverse damper amplitude. The losses are shown relative to the baseline (\(A_{\text{ex}} = \SI{0.0}{\micro\radian}\)) and error bars represent \SI{1}{\sigma}. Left: Measured (blue circles) and simulated (blue triangles) relative losses for the cycle with spill length of \SI{4.8}{\second}. Right: Measured (green circles) and simulated (green triangles) relative losses for the cycle with spill length of \SI{1.2}{\second}. For both panels, a hybrid model was used to estimate the losses: the shaped beam distribution was produced by the surrogate model and the slow-extraction and transport through the transfer lines were simulated with Xsuite.}
    \label{fig:sps_loss_vs_gain_s1}
\end{figure*}

For extraction measurements, the beam was sent to the transfer line and the aggregated signals from the beam-loss monitors (BLMs) surrounding each Lambertson septum, the profiles at the secondary-emission (SEM) grids, and the intensities at the start of the transfer line and at the target locations were recorded. The damper-configuration for the profile measurements was used again, except that the initial and final exciter tunes were widened slightly to \(0.395\) and \(0.445\), respectively, as a precaution against a slightly different chromaticity at extraction. The wider chirp range reduces the effective excitation but ensures that the full vertical tune spread of the beam is covered. The measurements were repeated for both spill lengths. For each cycle, the losses at the first Lambertson septum -- normalised to the reference case with the transverse damper switched off -- are shown as a function of \(A_{\text{ex}}\) in Fig.~\ref{fig:sps_loss_vs_gain_s1}, together with the relative losses predicted by the surrogate model. The losses decrease with increasing \(A_{\text{ex}}\): at the largest amplitude applied (\(A_{\text{ex}}=\SI{0.2158}{\micro\radian}\)) the loss reduction reaches \SI{23.5}{\percent} for the short spill. For the long spill, the largest loss reduction of \SI{22.5}{\percent} was already achieved at \(A_{\text{ex}}=\SI{0.1619}{\micro\radian}\). The model predictions track the measurements well for the long spill but diverge at large amplitudes for the short spill. The origin of this divergence is not conclusively established, we tentatively note that the transverse damper was known to be experiencing operational issues during this measurement campaign, which may have degraded its effective performance. Extrapolating the simulations to the hardware limit of the transverse damper amplitude \(A_{\text{ex}}=\SI{0.2698}{\micro\radian}\) yields a maximum loss reduction of \SI{62}{\percent} at the first Lambertson septum (see Fig.~\ref{fig:equal_sharing_losses}) and fewer than \SI{0.01}{\percent} of particles lost at other transfer line apertures.

\section{Amplitude detuning effects}\label{sect:ps}
\subsection{SPS simulations}
\begin{figure}
    \centering
    \includegraphics[width=1.0\columnwidth]{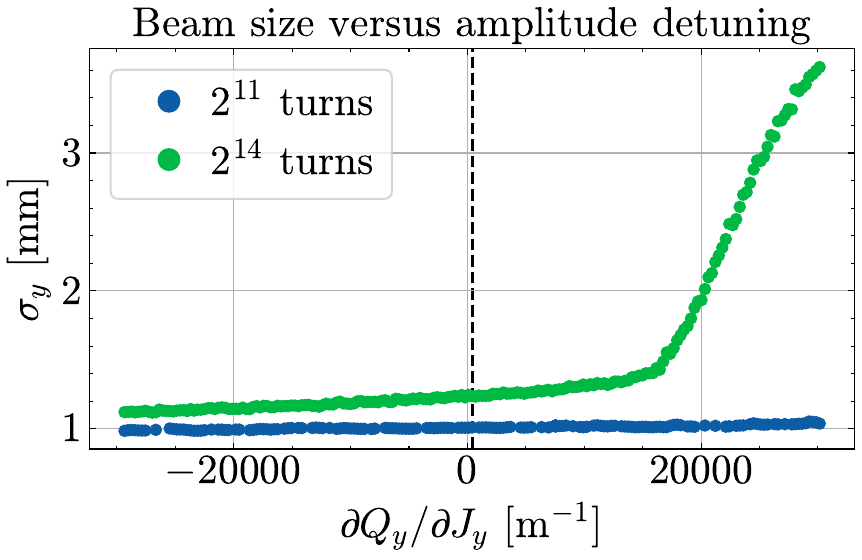}
    \caption{Vertical beam size \(\sigma_y\) versus amplitude detuning \(\partial Q_y / \partial J_y\). Results for two different configurations of the transverse damper are shown: \(A_{\text{ex}}\) = \SI{0.2158}{\micro\radian}, \(n=2^{11}\), \(q_{\text{ex, initial}} = 0.40\) and \(q_{\text{ex, final}} = 0.44\) (blue); \(A_{\text{ex}}\) = \SI{0.1}{\micro\radian}, \(n=2^{14}\), \(q_{\text{ex, initial}} = 0.40\) and \(q_{\text{ex, final}} = 0.44\) (green). The amplitude detuning of the SPS with the octupoles turned off is shown as black dashed line.}
    \label{fig:sps_detuning}
\end{figure}

The SPS measurements of Sec.~\ref{sect:measurements} were conducted with the octupoles switched off. Amplitude detuning is nonetheless present due to field errors and higher-order effects of the sextupoles; the latter account for \(\partial Q_y / \partial J_y =\SI[per-mode=power]{400.17}{\per\meter}\) according to the model. To assess the impact of \(\partial Q_y / \partial J_y\) on the excitation procedure, we scanned the normalised strength of a family of defocusing octupoles \(k_3 \in [\SI[per-mode=power]{-20}{\per\meter\tothe{4}}, \SI[per-mode=power]{20}{\per\meter\tothe{4}}]\) in the SPS model. For the chirp used in the measurements (\(A_{\text{ex}}=\SI{0.2158}{\micro\radian}\), \(n=2^{11}\), \(q_{\text{ex, initial}}=0.40\), \(q_{\text{ex, final}}=0.44\)) the effect is negligible (Fig.~\ref{fig:sps_detuning}). For a longer, weaker chirp (\(A_{\text{ex}}=\SI{0.1}{\micro\radian}\), \(n=2^{14}\)) the amplitude detuning significantly impacts the resulting beam profile.

\subsection{PS setup and measurements}\label{sect:ps_setup_measurements}
It was decided to measure the impact of the vertical amplitude detuning \(\partial Q_y / \partial J_y\) on the final distributions generated by the excitation in the PS instead of the SPS. This decision was taken for two reasons: first, the results of this study can be used to improve the PS model (see Sec.~\ref{sect:sys_id}), which benefits other studies, for instance that in Ref.~\cite{remta:ipac25-tupb018}. Second, beam time for machine development studies is scarcer in the SPS than in the PS. 

For this study, we used a single-bunch proton beam accelerated to \SI{20.32}{\giga\electronvolt\per\lightspeed} and de-bunched afterwards, as well as the three octupole families of the PS -- two identical focusing families with normalised strengths \(k_{3,1}\) and \(k_{3,2}\), and a defocusing family with strength \(k_{3,3}\). We scanned the octupole strengths by choosing reference values \(k^{\text{cmd}}_{3,1}=k^{\text{cmd}}_{3,2}=40\) and \(k^{\text{cmd}}_{3,3}=12\) and scaling them jointly by a common factor between \(-1.0\) and \(0.1\) to induce different \(\partial Q_y / \partial J_y\). These reference values were chosen to remain well within the operational limits of the magnets and the asymmetry between positive and negative scaling is due to losses increasing with positive normalised octupoles strengths. The transverse damper was configured with \(A^{\text{cmd}}_{\text{ex}}=\SI{1}{\micro\radian}\), \(n=2^{15}\), \(q_{\text{ex, initial}}=0.295\) and \(q_{\text{ex, final}}=0.325\). Note that the \textit{commanded} settings are stated here, which do not agree with the effective model obtained in Sec.~\ref{sect:sys_id}. The vertical beam profiles for these settings were measured with a wirescanner and are shown in Fig.~\ref{fig:ps_detuning_effects}(a): the more negative the scaling factor, the more the profile is distorted towards a bimodal distribution, and the extent of the beam tails shrinks.

\subsection{Effective model by system identification}\label{sect:sys_id}
\begin{figure}
    \centering
    \includegraphics[width=1.0\columnwidth]{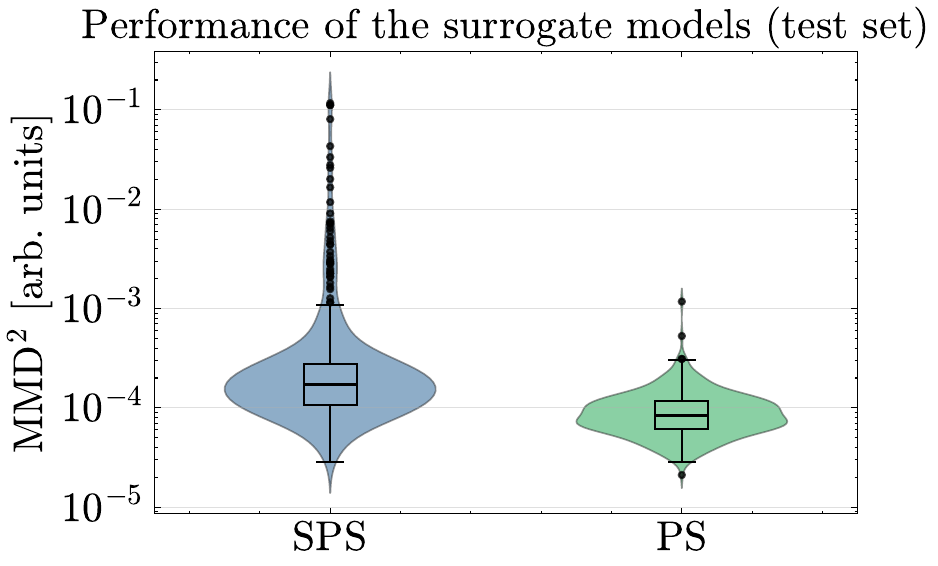}
    \caption{Box-plot and violin-plot of the \(\text{MMD}^2\)-performance achieved by the surrogate model for the SPS (blue) and the PS (green) on the respective test sets.}
    \label{fig:mmd_violin}
\end{figure}

We aim to compare the measured vertical beam profiles to single-particle tracking simulations, using a CFM surrogate as a fast intermediary. The PS-setup (see Sec.~\ref{sect:ps_setup_measurements}) was simulated with \texttt{Xsuite}, varying three parameters: the strength of the focusing families (\(k_{3,1}=k_{3,2}\)), the defocusing strength \(k_{3,3}\) and the transverse damper amplitude \(A_{\text{ex}}\). The bounds on each parameter are stated in Tab.~\ref{tab:param_bounds_ps}. We sampled \num{3424} points in this space, tracking \num{1e4} particles per point. A CFM surrogate model was trained on a subset of \num{2738} simulations, using the procedure described in Sec.~\ref{sect:surrogate_model}. The remaining \num{686} simulations were used for testing. The PS surrogate reaches substantially lower \(\text{MMD}^2\)-values than the SPS surrogate model (see Fig.~\ref{fig:mmd_violin}). This gap is consistent with the larger training set used for the PS case, though other factors (e.g.\ differences in the complexity of the distribution's response to parameter variation) cannot be fully excluded.

\begin{table}
\caption{Bounds of the parameter space for the PS simulations. 3424 Sobol points were sampled from this space.}
\label{tab:param_bounds_ps}
\begin{ruledtabular}
\begin{tabular}{ll}
Parameter & Range \\
\midrule
$k_{3,1}=k_{3,2}$ & $[\SI[per-mode=power]{-240}{\per\meter\tothe{4}},\ \SI[per-mode=power]{240}{\per\meter\tothe{4}}]$ \\
$k_{3,3}$ & $[\SI[per-mode=power]{-72}{\per\meter\tothe{4}},\ \SI[per-mode=power]{72}{\per\meter\tothe{4}}]$ \\
$A_{\text{ex}}$ & $[\SI{0}{\micro\radian},\ \SI{1}{\micro\radian}]$ \\
\end{tabular}
\end{ruledtabular}
\end{table}

\begin{figure*}
    \centering
    \includegraphics[width=1.0\textwidth]{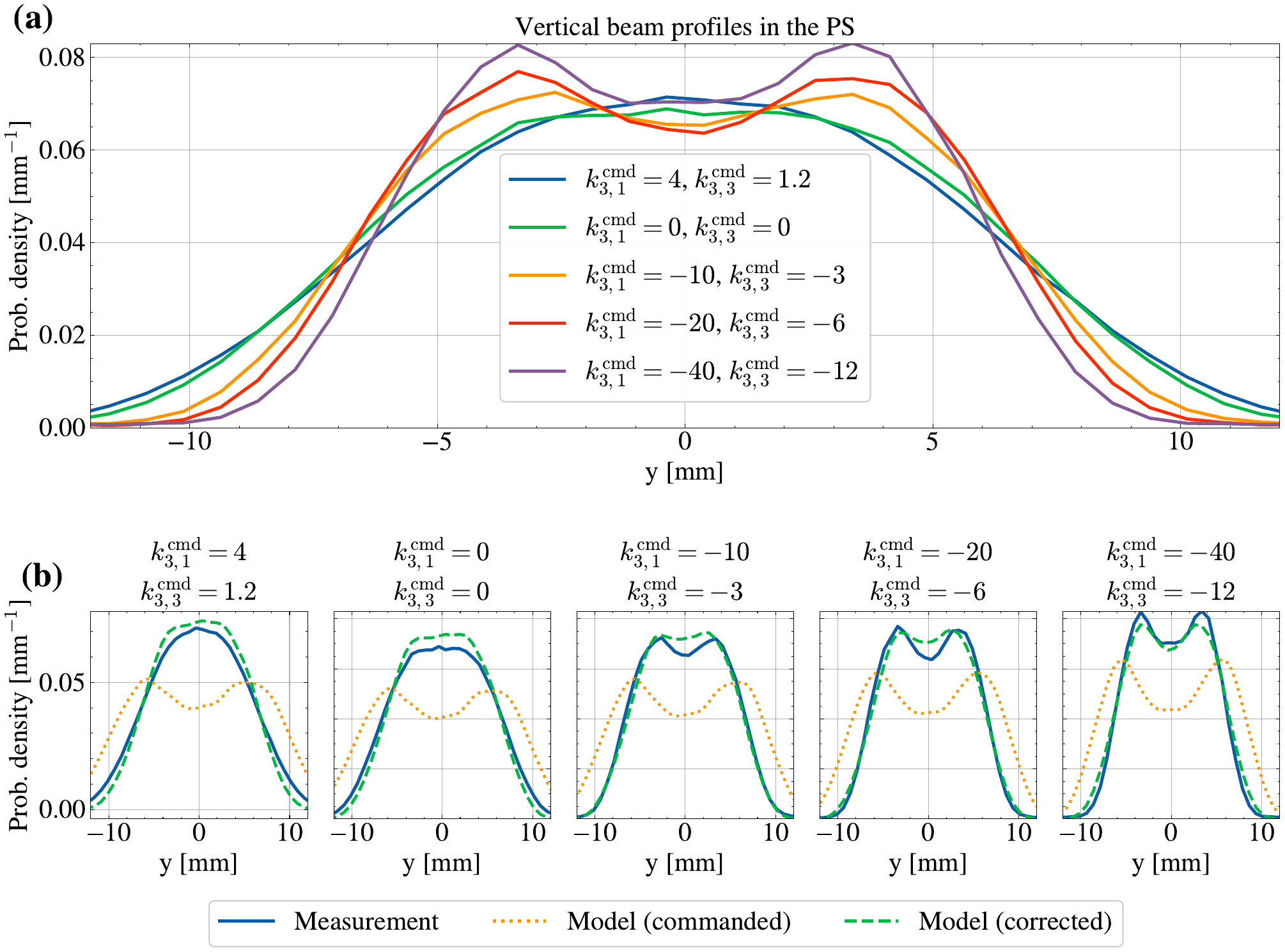}
    \caption{(a) Measured vertical beam profiles in the PS for a range of octupole
    settings, with the transverse exciter configured as \(A^{\text{cmd}}_{\text{ex}} =
    \SI{1}{\micro\radian}\), \(n = 2^{15}\) turns, and a tune chirp from
    \(q_{\text{ex, initial}} = 0.295\) to \(q_{\text{ex, final}} = 0.325\). (b) Measured profiles
    (blue) versus the surrogate-model prediction at the commanded strengths (orange dotted): a substantial mismatch persists even with the octupoles off. Scaling \(k_{3,3}\) by \num{1.84} (from the chromaticity measurement of Appendix~\ref{app:chroma}) with the fitted bias \(b = -5.1\), and \(A_{\text{ex}}\) by \num{0.62}, yields the corrected model (green dashed).}
    \label{fig:ps_detuning_effects}
\end{figure*}

Evaluated at the \emph{commanded} damper amplitude and octupole strengths, the surrogate reproduces the bimodal distortion qualitatively but shows a significant systematic mismatch at every setting (see Fig.~\ref{fig:ps_detuning_effects}(b)). Both commanded inputs are uncertain: the damper's gain-to-kick conversion is approximated from analytical formulas~\cite{Bass:2023}, and the machine's non-linear fields -- originating from magnet imperfections and the pole-face windings~\cite{juchno:ipac11-wepo019} -- are estimated from second-order chromaticity measurements (see Appendix~\ref{app:chroma} and Ref.~\cite{remta:ipac25-tupb018}). Since the PS has no vertical kicker, \(\partial Q_y / \partial J_y\) cannot be measured directly. Instead, we leverage Bayesian system identification to obtain an effective model: let \((m_i)_{i=1}^5\) be the profile measurements taken at commanded accelerator settings \((c_i^{\text{cmd}})_{i=1}^5\). We replace the commanded settings by corrected settings \(c_i^{\text{act}} = \phi(c_i^{\text{cmd}};\,\theta)\), where \(\phi\) is a parametric correction governed by parameters \(\theta\), and infer \(\theta\) such that the model evaluated at \(c_i^{\text{act}}\) reproduces \(m_i\).
Concretely, we perform an affine transformation of the commanded settings:
\begin{equation}
  A^{\text{act}}_{\text{ex}} = h\,A^{\text{cmd}}_{\text{ex}} , \hspace{0.5cm}
  k_3^{\text{act}} = g\,k_3^{\text{cmd}} + b,
  \label{eq:affine}
\end{equation}
with an amplitude gain \(h\), an octupole gain \(g\) capturing octupole calibration errors, and a bias \(b\) -- an effective octupole strength capturing detuning from other sources. We place uniform priors on these parameters, \(h\sim\mathcal{U}(0.01,1)\), \(g\sim\mathcal{U}(0,8)\) and \(b\sim\mathcal{U}(-10,10)\), and assign zero likelihood to corrected settings outside the surrogate's training range. The residual between measured and predicted profiles is modelled as multivariate Gaussian. Its covariance combines the shot-to-shot scatter of the measured profiles, the Monte Carlo noise of the finite surrogate sample, and a Gaussian-process model discrepancy that absorbs smooth systematic mismatch between simulation and machine~\cite{Kennedy2001}. We sample the posterior by Markov chain Monte Carlo, using the affine-invariant ensemble sampler~\cite{goodman2010ensemble} implemented in \texttt{emcee}~\cite{Foreman_Mackey_2013, Foreman-Mackey2019}.

\begin{figure}
  \centering
  \includegraphics[width=1.0\columnwidth]{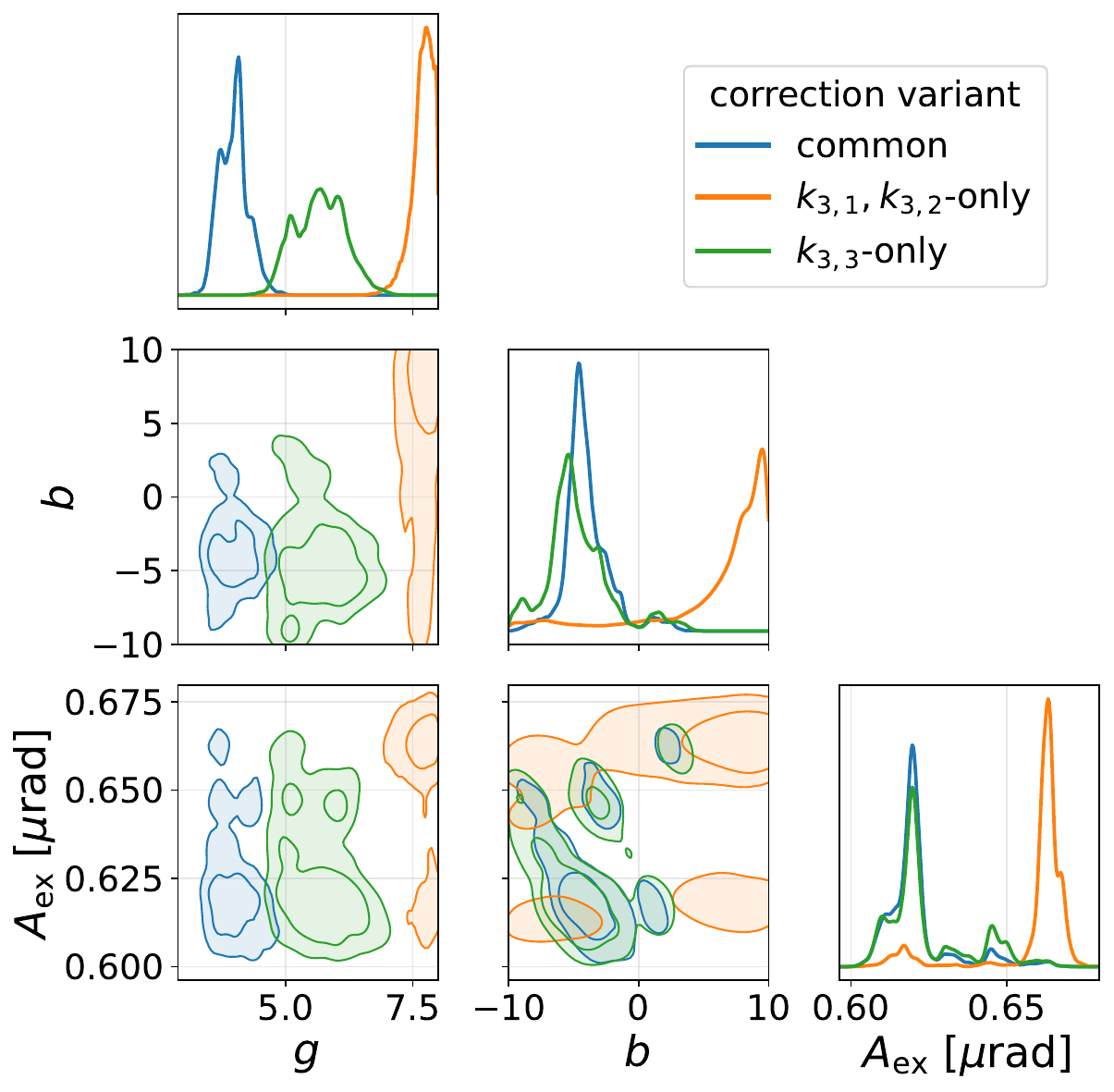}
  \caption{Joint posterior of the correction parameters \((g,\,b,\,A_{\text{ex}})\)
  for the three variants (one- and two-dimensional marginals, 68\% and 95\% contours). The gain is nearly uncorrelated with \(b\) and \(A_{\text{ex}}\). The \(k_{3,1},k_{3,2}\) variant rails against the prior bounds in \(g\) and \(b\), reflecting its inability to reproduce the measurements.}
  \label{fig:sysid_posterior}
\end{figure}

We consider three variants of the octupole correction: transforming all three families jointly, only the focusing families \(k_{3,1}\) and \(k_{3,2}\), or only the defocusing family \(k_{3,3}\). The posterior distributions of the correction parameters are shown for each variant in Fig.~\ref{fig:sysid_posterior}. Correcting all three families jointly and correcting only \(k_{3,3}\) yield gains well inside the prior, and the two variants fit the data equally well. The gain applied solely to the focusing families \(k_{3,1}\) and \(k_{3,2}\) rails against the upper limit of the prior, suggesting that this variant cannot explain the mismatch between measurements and simulations. The profiles alone do not distinguish the joint correction from the \(k_{3,3}\)-only correction, but second-order chromaticity measurements single out \(k_{3,3}\) (Appendix~\ref{app:chroma}). We therefore adopt the single-family correction on \(k_{3,3}\).

The effective exciter amplitude is \(A_{\text{ex}}\approx\SI{0.62}{\micro\radian}\) (\(h\approx 0.62\)), consistently below the commanded value in all variants (0.62--0.66), indicating that the damper delivers about two thirds of the nominal kick. The data favour a negative bias (\(b = -5.1\), \([-6.6,\,-2.8]\) 68\% credible interval), but \(b = 0\) lies within the 95\% credible interval. The posterior gain lies well above unity (\(g = 5.7\), \([5.1,\,6.2]\) 68\% credible interval), which at face value would imply an implausibly large octupole calibration error. An independent second-order chromaticity measurement (Appendix~\ref{app:chroma}, Ref.~\cite{remta:ipac25-tupb018}) gives a significantly smaller mismatch of \(g\approx 1.84\) for \(k_{3,3}\). Potentially, \(g\) absorbs other unmodelled effects, such as \(\beta\)-beating.

\section{Discussion}\label{sect:discussion}
The \SI{23.5}{\percent} loss reduction demonstrated here is not the intrinsic ceiling of the method. The chirp amplitude acts as a single continuous knob on the degree of deformation: the stronger the excitation, the larger the achievable loss reduction, until the dynamic aperture of the synchrotron or of the downstream transfer line becomes the limiting factor. At the maximum transverse damper amplitude, simulations give a \SI{62}{\percent} reduction with tolerable losses elsewhere in the transfer line, and a longer signal would extend this further. In practice, different sharing ratios and steering errors may shift this picture. A further limitation originates from the experiments: the NA primary targets are only \SI{2}{\milli\meter} high in the vertical plane~\cite{andreev2023measurementintrinsichadroniccontamination}, so the vertical beam size must remain small enough to stay on target. We did not probe the severity of this constraint here. The PS studies further show that the technique transfers between machines and is applicable across a wide range of momenta -- from \SIrange{20}{400}{\giga\electronvolt\per\lightspeed}. It could therefore be applied to other use cases, such as flattening the beam profile for fixed-target experiments~\cite{remta:ipac25-tupb018}. We note also that the agreement between surrogate and measurement is excellent for the long spill, but degraded for the short spill at large excitation amplitudes. Identifying the origin of this discrepancy will require dedicated follow-up measurements.

The vertical amplitude detuning \(\partial Q_y / \partial J_y\) partially alleviates these limitations by acting as a second, independent knob. Detuning of appropriate sign -- i.e.\ one that shifts the vertical tune in the opposite direction of the frequency sweep direction -- reduces the final beam envelope, as predicted by single-particle simulations and confirmed in the dedicated PS study. This requires sufficiently strong octupoles and can itself degrade the dynamic aperture. Crucially, the effect is negligible for the short, high-amplitude chirp used throughout the SPS measurements and becomes significant only for longer excitation signals, which extend the shaping time. This creates an operational trade-off: since the shaping is applied at the start of extraction, a short, strong chirp is preferable, because the beam must filament into the new steady-state distribution and a longer transient risks perturbing the splitting ratio -- and hence the spill flatness at the individual targets -- during this period.

Bayesian system identification combined with fast forward models is a promising approach for constructing effective machine models, and here provides a novel means of estimating \(\partial Q_y / \partial J_y\) without access to a vertical kicker. The identification cleanly recovers a correction to the transverse damper amplitude and additionally flags a calibration mismatch of the octupoles. The latter, however, must be interpreted with care: the inferred octupole gain is unphysically large -- a factor of about 3 above the value implied by second-order chromaticity measurements (\(g\approx 5.7\) versus \(g\approx 1.84\)) -- so that effects beyond the octupoles are likely folded into this single coefficient, causing the fit to overestimate the amplitude detuning. Plausible contributions are further model-to-machine discrepancies, e.g.\ \(\beta\)-beating, as well as measurement artefacts introduced by the wire scanner. The fidelity of the effective model is therefore limited by two factors: how completely the relevant physical effects are captured by the correction, and the quality of the underlying measurements.

\section{Conclusion}\label{sect:conclusion}
Vertical profile shaping with the transverse damper in the SPS offers a controllable and reproducible means of reducing splitting losses at the Lambertson septa in the NA transfer lines, and thereby the radioactive activation that constrains the annual proton flux to the NA~\cite{koto2022searchesnewphysicshighintensity}. A frequency-chirp excitation reshapes the Gaussian vertical profile into a uniform distribution with exponential tails and, at higher excitation amplitude, into a bimodal one -- in agreement with single-particle tracking and the surrogate model. In this study the splitting losses were reduced by up to \SI{23.5}{\percent}. This figure reflects the largest chirp amplitude tested rather than an intrinsic limit of the method. The same shaping-method was also successfully tested at the PS, showing that it is applicable at different machines and a wide range of energies.

Vertical amplitude detuning \(\partial Q_y/\partial J_y\) acts as a second, independent control: detuning of appropriate sign reduces the final beam envelope, as predicted in simulation and confirmed in the dedicated PS study, provided the excitation signal is long enough for the effect to be relevant.

Bayesian system identification with fast forward models is a promising route to effective machine models and, in the absence of a vertical kicker, allows \(\partial Q_y/\partial J_y\) to be estimated indirectly. Comparison with second-order chromaticity measurements shows, however, that the estimate absorbs additional unmodelled effects and is therefore substantially inflated; isolating and modelling those effects to recover a realistic value is left to future work.

As a next step, the profiles at the three primary targets should be measured and the transverse-damper gain adjusted such that the maximum loss reduction is achieved without compromising the yield of secondary particles.

\begin{acknowledgments}
M.R. designed and carried out the studies, developed the software, and wrote the manuscript. Y.D. and F.V. supervised the work, provided ideas and feedback in discussions, and contributed in the revision of the manuscript. M.R. used AI tools (Claude Opus 5 and Claude Opus 5.5, both Anthropic) to design and write code for the system identification study in Sec.~\ref{sect:sys_id}. Study protocols were drafted and approved by M.R. execution, results were carefully checked and replication pipelines are published in the companion repository. The aforementioned tools were also used to revise the manuscript. The authors take full responsibility for the content.
\end{acknowledgments}

\section*{Data and code availability}
Trained model checkpoints, simulation datasets and measurement data~\cite{datasets2026} used in this work are publicly available on Zenodo. Code is available on GitLab (\hyperlink{https://gitlab.cern.ch/abt-optics-and-code-repository/projects/ps-octupole-sysid.git}{https://gitlab.cern.ch/abt-optics-and-code-repository/projects/ps-octupole-sysid.git}).

\appendix
\section{Impact of beam sharing on losses}\label{app:sharing}
\begin{figure}
    \centering
    \includegraphics[width=1.0\columnwidth]{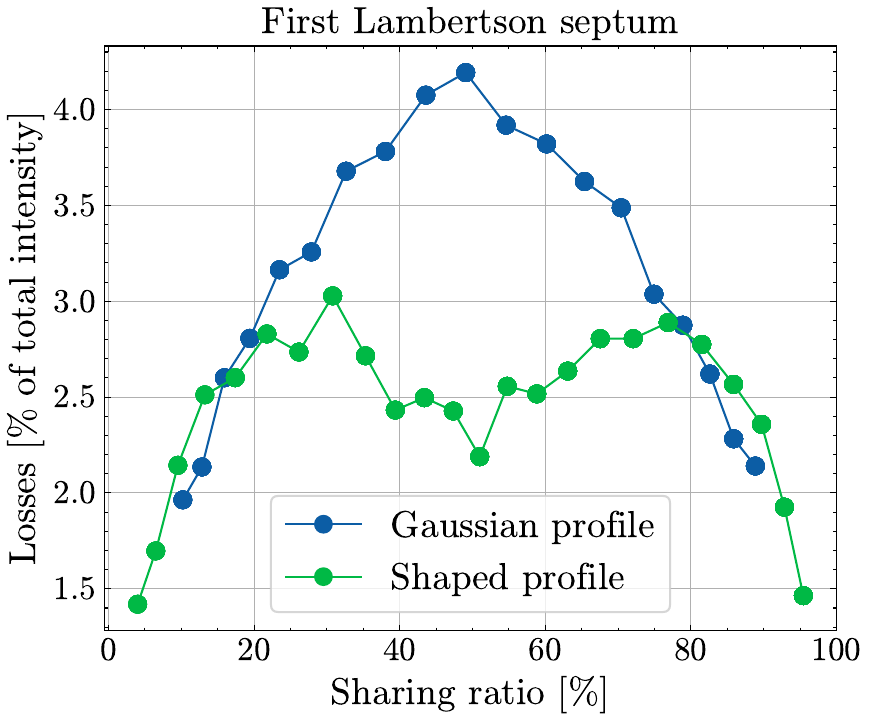}
    \caption{Simulated fraction of the total beam intensity lost versus sharing ratio at the first Lambertson septum for the baseline Gaussian profile (blue) and the shaped profile (green). The shaped profile was obtained with \(A_{\text{ex}}\) = \SI{0.2158}{\micro\radian}, \(n=2^{11}\), \(q_{\text{ex, initial}} = 0.40\) and \(q_{\text{ex, final}} = 0.44\) and linearly scaled to match the envelope of the Gaussian beam.}
    \label{fig:sharing_losses}
\end{figure}

\begin{figure}
    \centering
    \includegraphics[width=1.0\columnwidth]{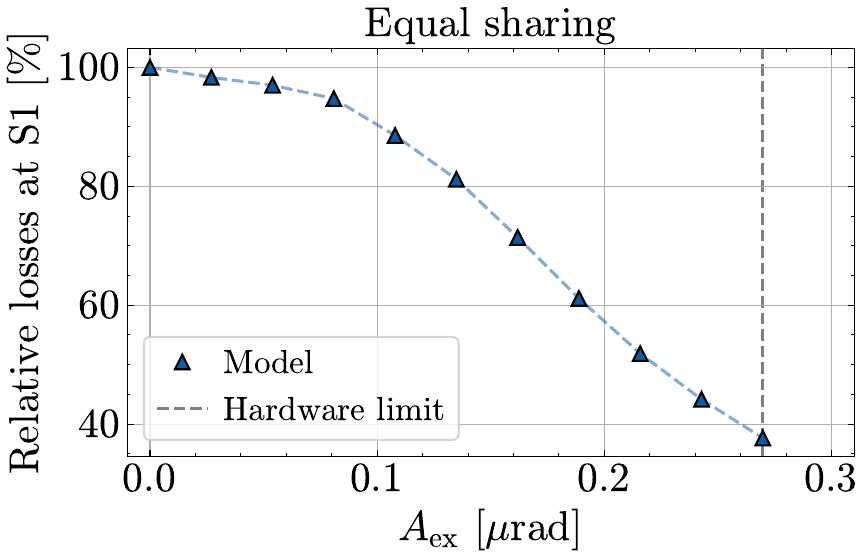}
    \caption{Simulated losses versus transverse damper amplitude \(A_{\text{ex}}\) at the first Lambertson septum, assuming an equal sharing ratio. The losses are shown relative to the baseline (\(A_{\text{ex}}\) = \SI{0.0}{\micro\radian}).  The other settings of the transverse damper were kept fixed at \(n=2^{11}\), \(q_{\text{ex, initial}} = 0.395\) and \(q_{\text{ex, final}} = 0.445\).}
    \label{fig:equal_sharing_losses}
\end{figure}

In general, the total intensity of the primary beam from the SPS is not shared equally between the three primary NA targets and the sharing ratio may be varied throughout operation by altering the vertical steering at the Lambertson septa. Consequently, the sharing ratio influences beam losses at the septa, because differently dense regions of the beam interact with the material. Figure~\ref{fig:sharing_losses} shows the simulated fraction of the total beam intensity lost as a function of the sharing ratio, both for the baseline Gaussian profile and the shaped beam distribution (\(A_{\text{ex}} =\) \SI{0.2158}{\micro\radian}). With a Gaussian beam profile, the highest losses -- approximately \SI{4}{\percent} of the total intensity -- are occurring with an equal sharing ratio. Conversely, this is the best case for the shaped distribution, with losses around \SI{2}{\percent} of the total intensity. The loss reduction versus transverse damper amplitude with an equal sharing ratio is shown in Fig.~\ref{fig:equal_sharing_losses}. At extreme sharing ratios, i.e. above \SI{80}{\percent}, the two distributions cause roughly equal losses.

\section{Second-order chromaticity measurements}\label{app:chroma}
\begin{figure}
    \centering
    \includegraphics[width=1.0\columnwidth]{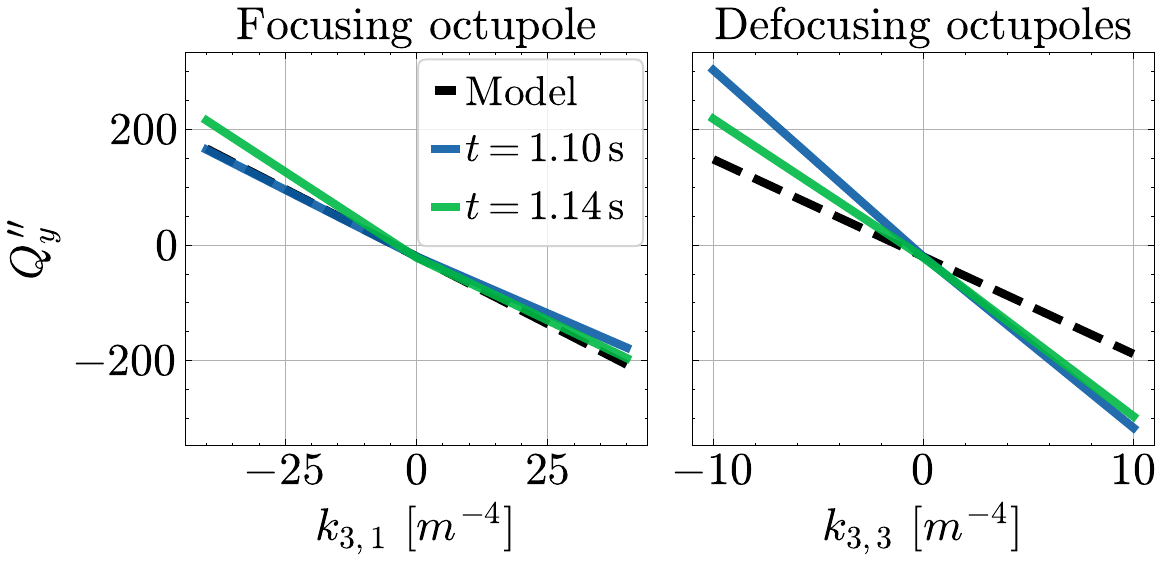}
    \caption{Measured vertical second-order chromaticity at two different times in the flat-top  of the cycle, namely at \SI{1.1}{\second} (blue) and \SI{1.14}{\second} (green) and the prediction of the uncorrected model (black dashed). Left: measured vertical second-order chromaticity for different commanded normalised strengths \(k_{3, 1}\). Right: measured vertical second-order chromaticity for different commanded normalised strengths \(k_{3, 3}\).}
    \label{fig:ps_chromaticity}
\end{figure}

Octupoles are the primary handle for correcting the second-order chromaticity \(Q^{\prime\prime}\)~\cite{1591526}. Conversely, we exploit repeated second-order chromaticity measurements at different commanded strengths to characterise the octupoles in the PS. Each measurement is performed by inducing coherent momentum offsets of varying amplitude and recording the resulting tune; a parabolic fit to these data yields the chromaticity coefficients up to second order. The results are shown in Fig.~\ref{fig:ps_chromaticity}. The measured and predicted \(Q^{\prime\prime}_y\) agree very well for \(k_{3,1}\) but differ significantly for \(k_{3,3}\). Comparing the measured slope of \(Q^{\prime\prime}_y\) versus commanded strength against the model prediction yields a correction factor for the model: \(k_{3,1}\) requires essentially no correction, whereas \(k_{3,3}\) is found to be a factor of \numrange{1.53}{1.84} stronger than predicted across repeated measurements. The remaining focusing family \(k_{3,2}\) is nominally identical to \(k_{3,1}\) and was therefore not measured separately. The value \(g=1.84\) used for the corrected-model overlay in Fig.~\ref{fig:ps_detuning_effects} corresponds to the upper end of the range measured for \(k_{3,3}\).

\FloatBarrier
\bibliographystyle{abbrvnat}
\bibliography{References}% Produces the bibliography via BibTeX.

\end{document}